# Impedance-tuned microwave loop for fast, homogeneous Rabi oscillations of a dense ensemble of NV-diamond electronic spins

*Han Sae Jung[1,‡], Johannes Cremer[2,‡], Aoyang Zhang[1,3,‡], Sangha Kim[1], Guang Yang[1], Ronald L. Walsworth[2,4,5,6,*], and Donhee Ham[1,*]*

[1]John A. Paulson School of Engineering and Applied Sciences, Harvard University, Cambridge, MA 02138, USA.

[2]Quantum Technology Center, University of Maryland, College Park, MD 20742, USA

[3]School of Integrated Circuits, Tsinghua University, Haidian District, Beijing 100084, China

[4]Department of Physics, University of Maryland, College Park, MD 20742, USA

[5]Joint Quantum Institute, University of Maryland, College Park, MD 20742, USA

[6]Department of Electrical and Computer Engineering, University of Maryland, College Park, MD 20742, USA

[‡]These authors contributed equally to this work.

*Corresponding author. Email: donhee@seas.harvard.edu, walsworth@umd.edu

ABSTRACT

Obtaining a high Rabi oscillation frequency homogeneously across a spatially-extended population of nitrogen-vacancy (NV) center electronic spins in diamond is useful for efficient spin-state manipulation of the NV ensemble and in using NVs to detect ensembles of other spin species. Here, we achieve a high, homogeneous Rabi frequency for a dense NV ensemble by enhancing the microwave magnetic fields in the center region of a diamond-coupled planar metallic loop via systematic engineering that increases the microwave current driving of the loop, while avoiding off-center proximity to the loop that gives strong but inhomogeneous microwave fields. With such enhanced microwave fields at 2.55 GHz, we achieve a 136.3 MHz NV Rabi frequency with 1.5% inhomogeneity over a 40 × 40 μm$^2$ diamond area; and use the NV ensemble to detect a ~30-MHz magnetic signal, similar to a nuclear magnetic resonance signal at a tesla-scale bias magnetic field, with Hz-scale spectral resolution.

TEXT

Electronic spin dynamics in diamond nitrogen-vacancy (NV) centers can have sufficient coherence (1,2) to exhibit Rabi oscillations when driven by a GHz microwave magnetic field tuned to the NV electron spin resonance (ESR) frequency $f_0$. (2,3) The ability to increase the NV Rabi oscillation frequency $f_1$ is often desired, because a larger $f_1$ for a given electronic spin dephasing time increases the number of quantum-coherent manipulations. (3) Since $f_1 = \gamma B_1/2$ where $B_1$ is the amplitude of the linearly driving microwave magnetic field and $\gamma$ is the NV electronic spin gyromagnetic ratio, (3) the problem of obtaining a larger Rabi frequency $f_1$ translates to a problem of driving a conducting wire or antenna with a larger amplitude GHz current to produce a larger $B_1$. Increasing a GHz driving field is typically not an easy problem in microwave engineering. (3–7)

Certain applications with NV electronic spins do not just desire a high $f_1$ but impose a requirement for a minimum $f_1$. A case in point is measurement scheme for nuclear magnetic resonance (NMR) signals—nuclear spin precession motions—using NV electronic spin dynamics as an NMR signal detector, or 'NV-NMR' for short. (8–12) In this scheme, in particular the one based on a class of pulsed dynamical decoupling protocols, NV electronic spins can detect frequency and phase information of nuclear spin precessions via dipole-dipole coupling, with the NV electronic spin dynamics read out optically. A key requirement for such NV-NMR measurements is that $f_1$ is significantly larger than the NMR frequency, *i.e.*, the nuclear spin precession frequency. NV-NMR can attain higher sensitivity for small samples than traditional NMR methods based on inductive readout, and thus can enable NMR spectroscopy for nanoscale and micron-scale samples. (8–12) At the same time, NV-NMR has its own challenges due to the requirement that $f_1$ be larger than the NMR frequency, which becomes especially pronounced as one seeks to increase the static magnetic field $B_0$ for higher-resolution NMR. As $B_0$ is increased, both $f_0$ and the NMR frequency increase, and so does the minimum required $f_1$. To meet this requirement, one has to drive a conducting wire with a larger current amplitude at a higher GHz frequency ($f_0$), working against the general trend that microwave driving becomes harder at a higher frequency. Due to this difficulty, high-resolution NMR readily achieved with the traditional inductive readout—*e.g.*, commercially available 300 ~ 600 MHz $^1$H NMR with $B_0 \approx 7.1 \sim 14.1$ T—is far beyond the current reach of NV-NMR.

Thus, whether for NV-NMR or for other applications, a wealth of research efforts have been committed to increasing $f_1$ for NV electronic spins. (3–7) In particular, large $B_1$ can be readily attained at GHz ESR frequencies for a loop of wire, even without excessive a current, at positions proximate to the wire as opposed to the center of the loop: as an extreme example, using such edge fields, $f_1 = 0.44$ GHz for $f_0 = 0.49$ GHz was obtained ($f_1/f_0 \sim 0.90$), albeit with significant $B_1$ inhomogeneity. (3) Such edge regions are still useful for applications involving a small number of NVs in a small spatial region (<1 μm), but is not suitable for applications relying on a spatially more distributed NV ensemble (>1 μm), such as used for micron-scale NV-NMR. (8–12) For such applications, which demand greater $B_1$ and $f_1$ homogeneity, the center region of the loop can be utilized; because $B_1$ is minimum at the center and increases towards the loop on the disk plane enclosed by the loop, it is maximally homogenous around the center. At the same time, given that $B_1$ is minimum at the center, obtaining a high target $B_1$ or $f_1$ is more difficult at and around the loop center, requiring an ability to drive in a larger-amplitude microwave current in the loop. For example, Ref. 4 achieved a millimeter-scale $B_1$ inhomogeneity less than 5% (normalized standard deviation) at $f_0 = 2.87$ GHz in the center region of a loop, but $f_1$ was only 14.3 MHz with $f_1/f_0 \sim 5.0 \times 10^{-3}$.

Here we report strong yet homogeneous quantum-coherent manipulation at GHz ESR frequencies of a large ensemble of NV electronic spins, which we obtain via systematic microwave engineering that increases the microwave current driving of a 3-turn planar loop coupled to a diamond (the "device"). Over a 40 μm × 40 μm diamond area in the center region of the loop, we achieve an $f_1$ as large as 136.3 MHz at $f_0 = 2.55$ GHz ($f_1/f_0 \sim 5.3 \times 10^{-2}$) with $B_1$ inhomogeneity about 1.5% (normalized standard deviation). This result further advances our long-term aim of high-resolution NV-NMR spectroscopy at tesla-scale bias magnetic fields, as the 40 μm × 40 μm area is on a par with the analyte size of NV-NMR targets. (10–12) Moreover, we demonstrate NV ensemble based detection of a ~30 MHz magnetic signal (similar to an NMR signal at ~0.7 T bias field) using the device.

Our device consists of a thin slab of diamond (2.0 mm × 2.0 mm × 0.5 mm) glued on a chip (Fig. 1a, right), which features the 3-turn Cu loop supported by a ~8-μm-thick dielectric atop a 75-μm thick semi-insulating SiC substrate. The surface of the diamond slab immediately facing the loop embeds an ensemble of NV centers within a ~10-μm depth. The NV enriched diamond layer is grown using chemical vapor deposition (CVD) and contains 99.99% $^{12}C$ (~0.01% $^{13}C$) and 17 ppm $^{15}N$, yielding ~2.7 ppm $NV^-$ concentration. $T_2$ coherence times of the NV centers are 7.1 μs for a Hahn-Echo sequence and 33.0 μs for an XY8-6 sequence. The diamond slab covers the loop in its entirety. The diameter of the 3 Cu turns is 300 μm, 360 μm, and 420 μm, with an effective wire thickness of 3 μm, 9 μm, and 9 μm, respectively; the width of each Cu turn is 17 μm (Figure S1). The bottom of the chip below the SiC substrate is finished with a 10-μm-thick gold layer. The additional single loop surrounding, but not connected to the 3-turn loop, seen in Figure 1a, right, was designed as a secondary loop to shim the microwave field produced by the primary 3-turn loop to further reduce its inhomogeneity; however, mutual inductance between the secondary and primary loops proved significant enough to affect the primary coil driving, so the secondary loop is left open-circuited and unused.

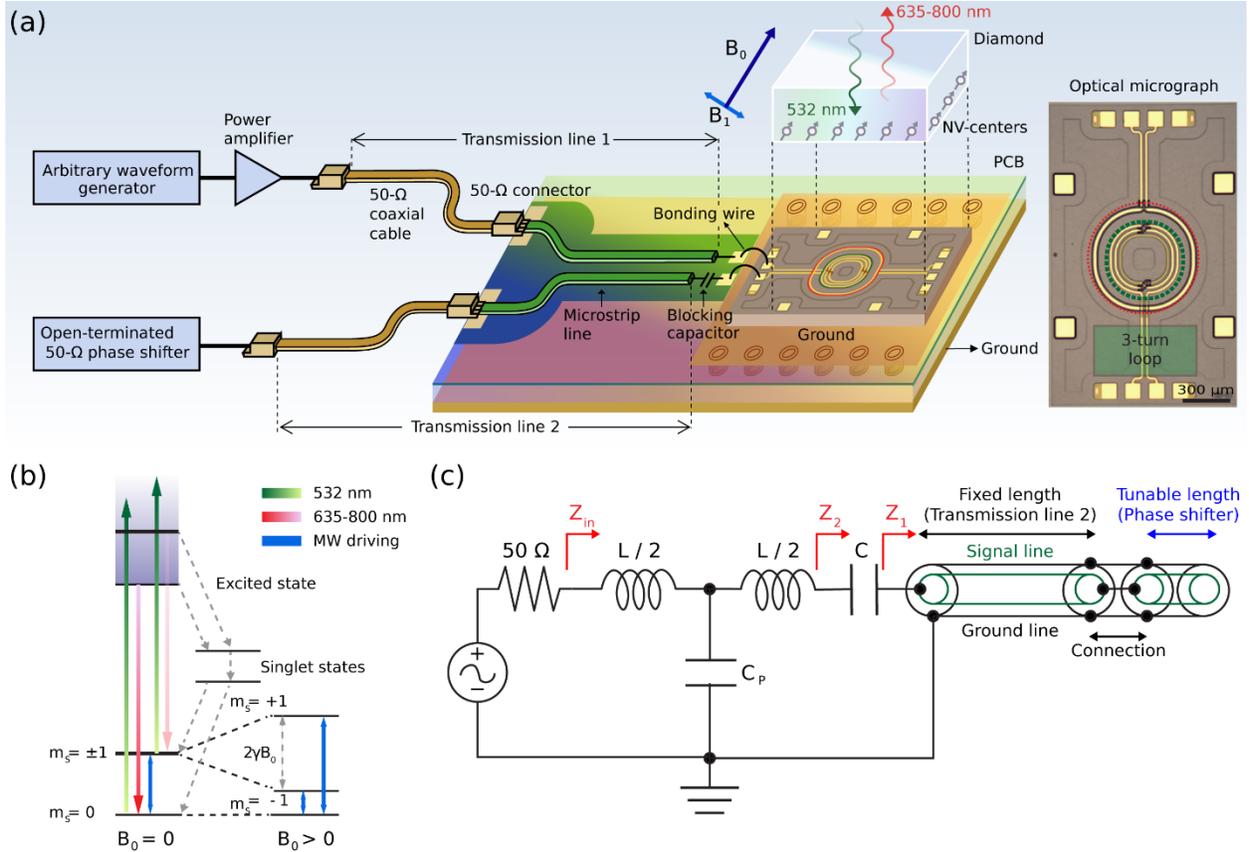

**Figure 1.** (a) Schematic of the device and measurement setup showing the full microwave signal pathway detailed in the main text (left) and an optical micrograph of the chip featuring the 3-turn planar Cu loop (right). The diamond slab is glued on top of the chip such that its NV-embedded surface layer (≈10 μm thick) faces the loop. The 3-turn loop, which is denoted by the dashed green circle in the optical micrograph (right), is the microwave antenna used in this work; whereas the additional single loop, denoted by the dashed red circle, is left open-circuited and unused. (b) Schematic of NV electronic spin states, optical and microwave transitions, and dependence on the static (bias) magnetic field, $B_0$. (c) Circuit model of the microwave signal pathway, with detail given in the main text. For maximal current delivery into the loop, the magnitude of the impedance $Z_{in}$ viewed by the microwave source (*i.e.*, the arbitrary waveform generator followed by the 50-Ω power amplifier) needs to be minimized, which is achieved by tuning the phase shifter.

When NV electronic spins are subject to a static (bias) magnetic field, $B_0$, two ESR frequencies arise: $f_0 = \Delta + \gamma B_0$ for the splitting between $m_s = 0$ and $m_s = +1$ states; and $f_0 = |\Delta - \gamma B_0|$ for the splitting between $m_s = 0$ and $m_s = -1$ states, with $\Delta \approx 2.87$ GHz due to the zero-field splitting (Figure 1b). The microwave drive frequency can be tuned to either of the two ESR frequencies in order to cause a Rabi oscillation between two correspondingly chosen NV spin states. Importantly, in our setup, the class of NV centers investigated in this study have their quantization axis (associated with one crystal axis within the diamond sample) tilted 54.7° away from the normal to the chip surface; and values of $B_0$ and $B_1$ given throughout this paper are all in reference to this NV-center axis. Concretely: the static $B_0$ field is the projection of the permanent-magnet-generated static magnetic field onto the relevant NV-center axis (hence by tuning the distance and orientation between the magnet and our device, we can tune $B_0$); and the microwave $B_1$ field is the projection of the loop-produced microwave field onto the plane perpendicular to the relevant NV-center axis. The optical initialization of the NV spins to the $m_s = 0$ state before microwave driving, and the optical readout of Rabi oscillations during microwave driving, follow the standard protocol (Figure 1b; Supporting Information). (10–12)

The full microwave signal path can be seen in Figure 1a. The chip with the diamond on top is mounted on a printed circuit board (PCB), which hosts two 50-Ω microstrip lines. One end of the 3-turn loop is wire-bonded to one of the two microstrips on the PCB, which connects to a microwave source—an arbitrary waveform generator followed by a power amplifier (ZHL-25W-63+) with a 50-Ω output impedance—via a 50-Ω coaxial cable. We call this pair of the microstrip and the coaxial cable 'Transmission line 1' (characteristic impedance: 50 Ω). The other end of the 3-turn loop is wire-bonded to a blocking capacitor $C = 0.5$ pF in series with the other microstrip on the PCB, which connects to an open-terminated 50-Ω phase shifter via another 50-Ω coaxial cable. We call this second pair of the microstrip line and the coaxial cable 'Transmission line 2' (characteristic impedance: 50 Ω). The gold backside of the chip is glued with conductive silver epoxy on a ground metal on the PCB, which is connected to the ground plane the two microstrip lines share.

Transmission line 1 delivers a signal from the microwave source to the loop to drive the NV electronic spins. The sequence of the blocking capacitor, Transmission line 2, and open-terminated phase shifter (which is a transmission line with an effective tunable length and thus a tunable phase delay $\phi$) combines to add an adequate impedance to the loop to maximize the injected microwave current and hence $f_1$. To see how this overall configuration works, consider the microwave signal path model shown in Figure 1c. $L$ models the inductance of the loop (Figure S2) and the two bonding wires together, whereas $C_p$ models the parasitic capacitance between the loop and the chip backside ground, which is the microstrips' ground ($C_p$ may be ignored to the first order approximation, given the substrate thickness). The 50-Ω microwave source matched to 50-Ω Transmission line 1 on the left of $L$ can be modeled as a voltage source with a 50-Ω source impedance. This source sees an impedance of $Z_{in}$ to the right, whose magnitude we seek to minimize (ideally to zero) to send a maximum microwave current into the loop (*i.e.*, the impedance is intentionally mismatched). Since 50-Ω Transmission line 2 with a fixed length (thus fixed phase delay $\phi_0$) together with the 50-Ω open-terminated phase shifter with the tunable phase delay $\phi$ presents a purely imaginary impedance of $Z_1(\phi) = -iZ_0 \times \cot(\phi + \phi_0)$ ($Z_0 \equiv 50$ Ω) to $C$, the loop sees, to its right, a purely imaginary impedance of $Z_2(\phi) = -i[Z_0 \cot(\phi + \phi_0) + (\omega C)^{-1}]$ ($\omega = 2\pi f$: angular microwave frequency). The imaginary part of $Z_1(\phi)$ is varied from $-\infty$ to $+\infty$ as $\phi$ is scanned with a period of $\pi$, and so is that of $Z_2(\phi)$. Then with a properly chosen $\phi$, $Z_2(\phi)$ can cancel the loop impedance of $i\omega L$ (ignoring $C_p$, to first order approximation) to achieve $Z_{in} \approx 0$, maximizing the microwave current delivery into the loop.

A few points merit discussion. First, variation of the imaginary part of $Z_1(\phi)$ from $-\infty$ to $+\infty$ is nominally sufficient, without the blocking capacitor that adds the $(\omega C)^{-1}$ term to the imaginary part of $Z_1(\phi)$, to tune $Z_{in}$ to zero. However, such tuning is based on the formula $Z_1(\phi) = -iZ_0 \times \cot(\phi + \phi_0)$ from the lossless model; in the actual case with loss, not only does $Z_1(\phi)$ have both real and imaginary parts, but its imaginary part has a finite range instead of $[-\infty, +\infty]$. Hence in the real device, the imaginary part of $Z_2(\phi)$, with the added term of $(\omega C)^{-1}$ from the blocking capacitor, has a range shifted from the imaginary part $Z_1(\phi)$, which helps to cancel the inductive impedance of $i\omega L$. Second, there is another view to this impedance engineering. To achieve $Z_{in} = 0$, the inductive impedance $i\omega L$ can be canceled simply by the blocking capacitor impedance, $-i(\omega C)^{-1}$, without adding the open-terminated phase shifter, if we choose a proper $C$ at a given frequency. However, in our experiment, we try various ESR frequencies by tuning $B_0$, but do not wish to keep changing the $C$ value to adapt to each frequency; thus we fix $C$ (0.5 pF), and seek to remove the remnant impedance of $i\omega L - i(\omega C)^{-1}$ by the tuning of $Z_1(\phi)$ via the open-terminated phase shifter. Third, not only loss but also various detailed parasitic elements are not included in the model, and the minimized $|Z_{in}|$ in practice is not zero: while this is less ideal from the point of view of our goal to maximize microwave current driving of the loop, it helps reduce the power reflection back to the source, by lowering the impedance mismatch between 50 $\Omega$ of the source and $Z_{in}$.

Figure 2a shows NV measurements of $f_1$ at the loop center vs. tuned phase delay $\phi$ (blue), juxtaposed with $f_1$ measured at the loop center with a fixed 50-$\Omega$ load replacing the phase-tunable open-terminated phase shifter (red, dashed line; see Figure S3 for its circuit diagram), where $f_0 =$ 2.55 GHz for the splitting between $m_s = 0$ and $m_s = -1$ ($B_0 = 116$ G in the present experiment). Measured values of $f_1$ vary widely with phase ($\phi$) tuning in the former case, going beyond and below the fixed $f_1$ value with fixed 50-$\Omega$ load termination in the latter case. This result demonstrates that with phase tuning, $Z_{in}$, and thus the microwave current delivered into the loop, vary widely. At $\phi \approx 100°$, $f_1$ is maximized at about 136.3 MHz with $f_0 = 2.55$ GHz, which is a notably fast NV Rabi oscillation given it is measured at the loop center (see Table S1 for comparison to the existing literature). For these conditions, Figures 2b and 2c show the NV optically detected magnetic resonance (ODMR) spectrum (i.e., the optically detected ESR spectrum) and an example Rabi oscillation, respectively; and Figure 2d shows the Fourier transform of the Rabi oscillation with a single peak at $f_1 = 136.3$ MHz (see Figure S4 for example ODMR and Rabi dynamics at different $f_0$ values). Figure 2e shows the maximum $f_1$ (Rabi frequency) at the center of the loop obtained by phase ($\phi$) tuning in comparison to measured $f_1$ at fixed 50-$\Omega$ load for a number of different $f_0$ values between 1.40 GHz and 6.02 GHz (varied by the aforementioned $B_0$ tuning). As in the case of $f_0 = 2.55$ GHz, the optimally tuned system drives a faster Rabi oscillation than the case of fixed 50-$\Omega$ termination. Likewise, the experimental driving efficiency, defined as the ratio of $f_1$ to the square root of the output power from the power amplifier, is also higher for the optimally tuned system for all $f_0$ values tested (Figure 2f; see also Figure S5).

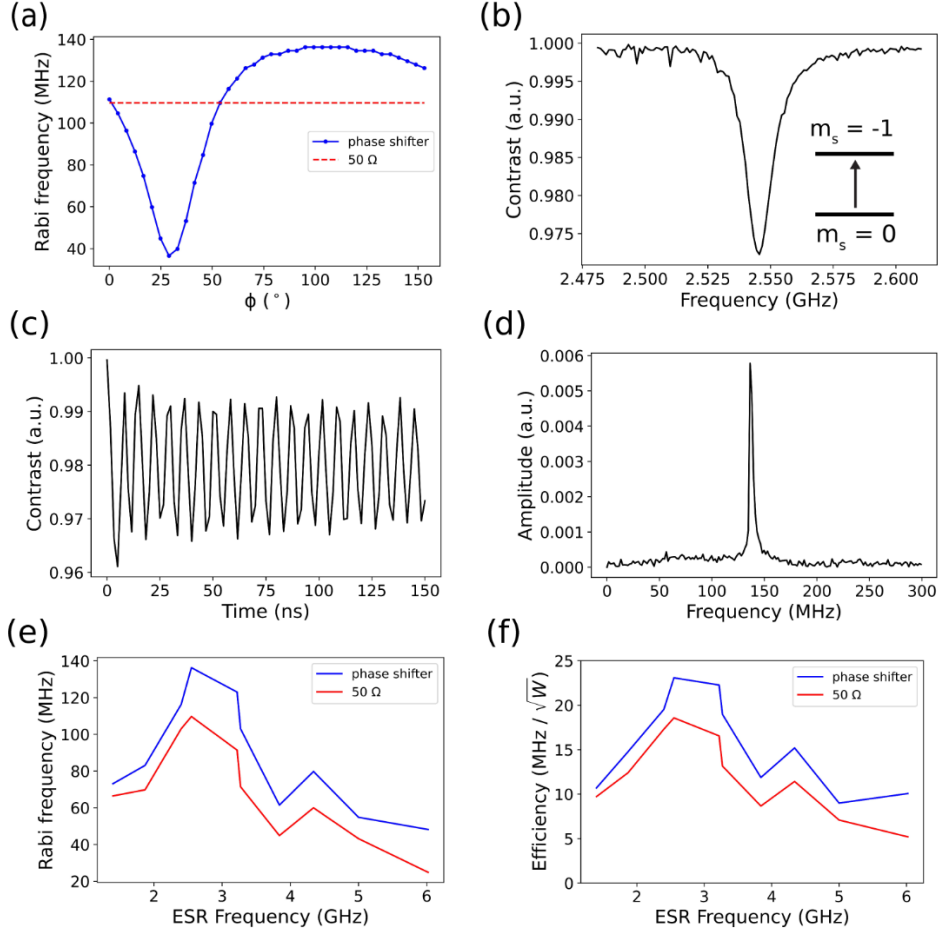

**Figure 2.** (a) NV measurements of $f_1$ at the center of the 3-turn loop as a function of phase delay $\phi$ of the open-terminated phase shifter (blue) and with fixed 50-$\Omega$ termination replacing the open-terminated phase shifter (red), where $f_0 = 2.55$ GHz for the splitting between $m_s = 0$ and $m_s = -1$ ($B_0 = 116$ G). At $\phi \approx 100°$, $f_1$ is maximized at 136.3 MHz. (b, c, d) Example measured NV ODMR spectrum, Rabi oscillation, and Fourier transform of the Rabi oscillation with a single peak at $f_1 = 136.3$ MHz, all for $\phi \approx 100°$ that yields the maximum $f_1$ at $f_0 = 2.55$ GHz. (e) Experimentally-determined maximum $f_1$ (Rabi frequency) at the center of the 3-turn loop obtained by phase ($\phi$) tuning (blue) in comparison to $f_1$ at fixed 50-$\Omega$ termination (red) for a number of different $f_0$ values (ESR frequencies) between 1.40 GHz to 6.02 GHz. (f) Repetition of (e) for experimental driving efficiency, defined as the ratio of $f_1$ to the square root of the output power from the power amplifier.

The optimally tuned 3-turn loop achieves both strong and homogeneous microwave $B_1$ at the loop center. Figure 3a, left and right, show $f_1$-maps (these may be considered as $B_1$-maps as $f_1 \propto B_1$) measured with the NV ensemble across a 280 μm × 280 μm area surrounding the innermost turn of the 3-turn loop, at $f_0$ = 2.55 GHz and $f_0$ = 6.02 GHz, respectively, using a scanned 5-μm diameter laser spot and a 10-μm pixel pitch. These two maps are each made with an optimally tuned ϕ value that yields the maximal $f_1$ values at the center, as determined from the blue curve at 2.55 GHz and 6.02 GHz in Figure 2e. Within a 40 μm × 40 μm area around the center of the 3-turn loop, $B_1$ does not exhibit appreciable inhomogeneity, and measured NV Rabi oscillations at each point in this area are slowly damped, with a Fourier transform showing a single, sharp peak (Figures S6 and S7). The $B_1$ inhomogeneity becomes larger across a 100 μm × 100 μm area (Figure 3b), with Rabi oscillations more rapidly damped and the Fourier transform broadened at each measurement point in the outer region of the 100 μm × 100 μm area (Figures S6 and S7). Quantitatively, at $f_0$ = 2.55 GHz, the mean of $f_1$ is 139.3 MHz and 150.3 MHz, with normalized standard deviations of about 1.5% and 6.3% inside the 40 μm × 40 μm and 100 μm × 100 μm areas, respectively (Figure 3c, left). At $f_0$ = 6.02 GHz, the mean of $f_1$ is 48.9 MHz and 53.2 MHz, with normalized standard deviations of about 1.7% and 6.8% inside the 40 μm × 40 μm and 100 μm × 100 μm areas, respectively (Figure 3c, right). This demonstrated $B_1$ homogeneity within the 40 μm × 40 μm NV-diamond area is expected to be sufficient for high-resolution, tesla-scale NV-NMR spectroscopy of a picoliter-scale analyte (10–15).

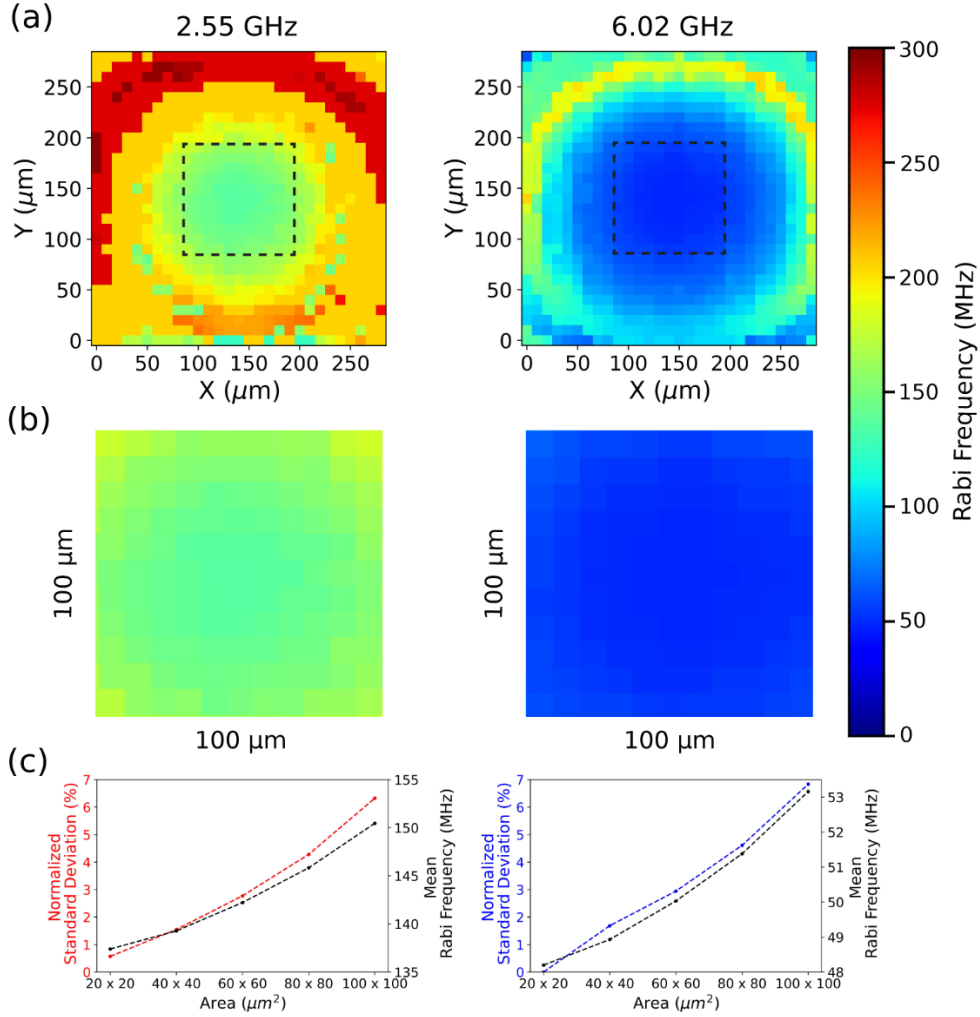

**Figure 3.** (a) NV-measured maps of $f_1$ across a 280 μm × 280 μm area surrounding the innermost turn of the 3-turn loop at $f_0$ = 2.55 GHz (left) and $f_0$ = 6.02 GHz (right). The two map are with specific, optimally tuned ϕ values that yield the maximal $f_1$ values at the center, corresponding to the blue curve values at 2.55 GHz and 6.02 GHz in Figure 2e. (b) Zoom in of $f_1$ maps from (a) for the 100 μm × 100 μm area around the loop center. This area is indicated by the dashed rectangles in (a). (c) Mean and normalized standard deviation of $f_1$ for varying areas around the loop center with $f_0$ = 2.55 GHz (left) and $f_0$ = 6.02 GHz (right), obtained from $f_1$ maps in (a). The normalized standard deviations are less than 2% inside the 40 μm × 40 μm area for both $f_0$ values.

Finally, we use our device to detect a 29.992 MHz radio frequency (RF) magnetic field signal generated by a test coil to emulate a tesla-scale NMR signal. The RF test signal is measured by the NV ensemble using coherently averaged synchronized readout (CASR), (10,11) a quantum spectroscopy protocol that alternates dynamical decoupling and optical readout of the NV electronic spins to provide Hz-scale spectral resolution (Figure 4a). We perform the measurements at $f_0 = 2.55$ GHz ($B_0 = 116$ G), with optimal phase ($\phi$) tuning yielding the maximum $f_1$ of 136.3 MHz, which is sufficiently large to enable phase-coherent measurements of the 29.992 MHz RF test signal. (8-12) The NV CASR measurement uses an XY8-6 sequence with a central frequency $f_{\text{casr}} \approx 30$ MHz. The CASR protocol effectively down-converts the RF signal frequency, and we observe an oscillatory variation in NV photoluminescence (PL) intensity at the down-converted frequency of $\delta f \approx 30$ MHz – 29.992 MHz = 8 kHz. The Fourier transform of observed PL time-domain signal (Figure 4b, left), measured for 1 second, shows a signal peak at 8.008 kHz (Figure 4b, right), with the 8 Hz deviation from 8 kHz being well less than the 1 ppm error of the RF signal generator (Rigol DG 1032). Increasing the CASR measurement time from 1 second to 10 seconds to 100 seconds, the spectral resolution of the RF test signal improves from about 1 Hz to 0.1 Hz to 50 mHz, eventually limited by drift in the system (Figure S8).

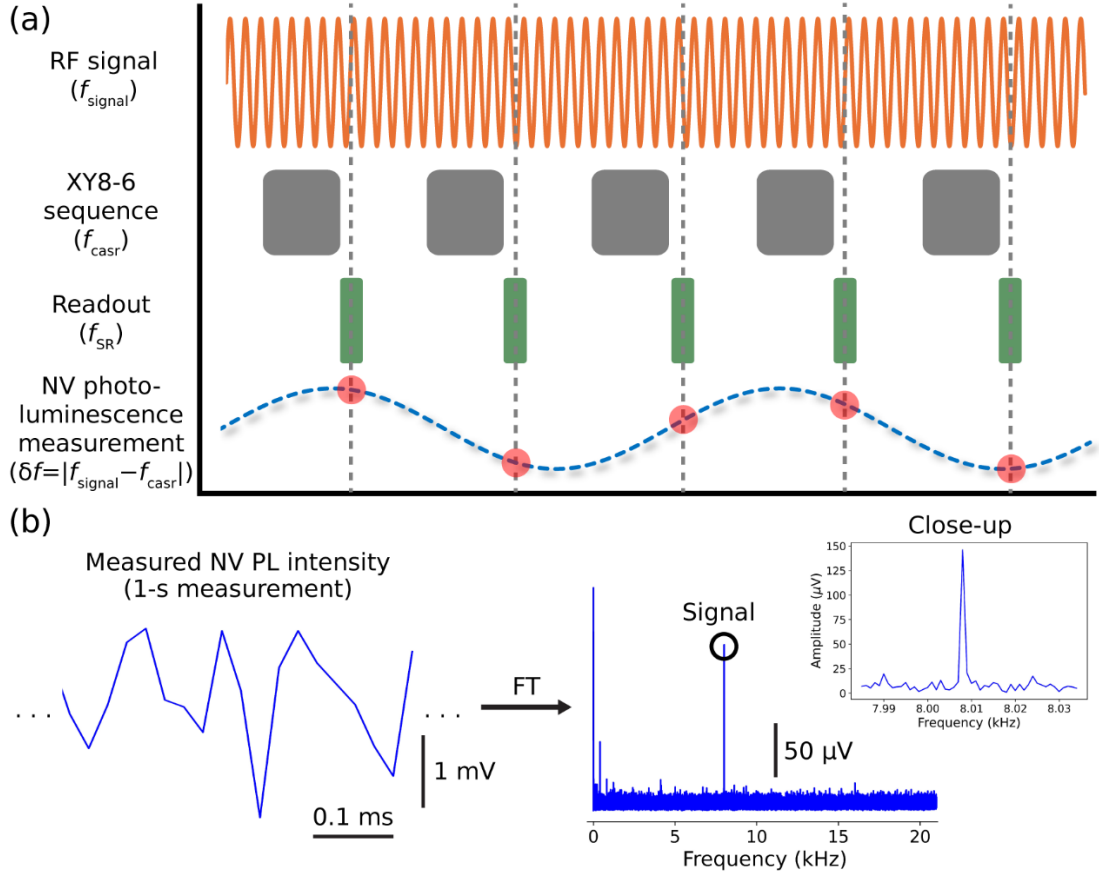

**Figure 4.** (a) CASR timing diagram for NV ensemble detection of a 29.992-MHz ($f_{signal}$) RF magnetic field from a test coil. NV electronic spins are driven at $f_0 = 2.55$ GHz ($B_0 = 116$ G)—with the maximal $f_1$ of 136.3 MHz obtained by phase delay ($\phi$) tuning—using alternating application of an XY8-6 dynamical decoupling sequence and NV photoluminescence (PL) readout. The CASR protocol produces an oscillatory variation in NV PL at frequency $\delta f = f_{signal} - f_{casr}$, where $f_{casr} \approx 30$ MHz is the central frequency used in the XY8-6 sequence. (b) Example measured NV PL intensity (left) and its Fourier transform (right), showing the signal peak at 8.008 kHz, in good agreement with expectations given the 1 ppm error of the signal generator used in this measurement.

This work demonstrates an approach to realize high Rabi frequency for a dense NV ensemble, with good spatial homogeneity, by increasing a microwave magnetic field in the center of a multi-turn wire loop with careful microwave engineering, instead of relying on proximity to the loop. Specifically, at an NV driving frequency of 2.55 GHz ($B_0$ = 116 G), we achieve a 136.3 MHz NV Rabi frequency with 1.5% inhomogeneity over a 40 × 40 μm$^2$ diamond area. We demonstrate the utility of our approach by using the NV ensemble in this diamond area to detect a ~30-MHz magnetic signal with Hz-scale spectral resolution. This capability may benefit high Rabi frequency applications that employ a micron-scale ensemble of NV electronic spins, such as tesla-scale NV-NMR, albeit significant additional engineering challenges remain for NV-NMR at such high bias magnetic fields. In envisioned future work, a driving amplifier will be custom designed with a microwave loop in the same chip and will be fabricated using a high-bandgap semiconductor (*e.g.*, GaN) technology, such that the microwave current is further increased due to both the short distance between the active and passive circuits and also their co-design.

ASSOCIATED CONTENT

**Supporting Information**.

- Methods, comparison of device used in present work to several other works in the literature, experimental setup, electromagnetic simulation of the 3-turn loop, circuit model of the microwave signal pathway using the 50-Ω termination, NV ODMR and Rabi oscillation measurements at various $f_0$ values, output power of the ZHL-25W-63+ power amplifier as a function of microwave excitation frequency, $f_1$ measurements at varying distances from the center of the 3-turn loop for the NV-measured maps of $f_1$ shown in Figure 3 at $f_0$ = 2.55 GHz and $f_0$ = 6.02 GHz, determination of the spectral resolution for the CASR measurement shown in Figure 4 at varying measurement times.

AUTHOR INFORMATION

**Corresponding Authors**

**Donhee Ham** - John A. Paulson School of Engineering and Applied Sciences, Harvard University, Cambridge, MA 02138, USA; Email: donhee@seas.harvard.edu

**Ronald L. Walsworth** - Quantum Technology Center, University of Maryland, College Park, MD 20742, USA; Department of Physics, University of Maryland, College Park, MD 20742, USA; Joint Quantum Institute, University of Maryland, College Park, MD 20742, USA; Department of Electrical and Computer Engineering, University of Maryland, College Park, MD 20742, USA; Email: walsworth@umd.edu

**Present Addresses**

Aoyang Zhang was with John A. Paulson School of Engineering and Applied Sciences, Harvard University, Cambridge, MA 02138, USA. He is now with the School of Integrated Circuits, Tsinghua University, Beijing 100084, China. Guang Yang was also with John A. Paulson School of Engineering and Applied Sciences, Harvard University, Cambridge, MA 02138, USA. He is now with the Anlyan Center, Yale University School of Medicine, New Haven, CT 06519, USA.

**Author Contributions**

H.S.J., J.C., and A.Z. contributed equally to this work. D.H. and R.L.W. conceived the work. H.S.J., J.C., and S.K. performed the experiments. A.Z. designed the 3-turn loop and the PCB. J.C., H.S.J., and A.Z. put together the measurement setup. H.S.J. and G.Y. packaged the chip. D.H. and R.L.W. supervised the project. H.S.J., J.C., A.Z., R.L.W., and D.H. wrote the manuscript, and all authors read and discussed it.

**Notes**

The authors declare no competing financial interest.

ACKNOWLEDGMENT

This work was supported by the Gordon and Betty Moore Foundation under Contract 7797.01, the Advanced Research Projects Agency-Energy (ARPA-E; Program Directors, Dr. Isik Kizilyalli and Dr. Olga Spahn) under Contract DE-AR0001063, the U.S. Army Research Laboratory under Contract No. W911NF2420143, and the University of Maryland Quantum Technology Center.

**Supporting Information:**

**Impedance-tuned microwave loop for fast, homogeneous Rabi oscillations of a dense ensemble of NV-diamond electronic spins**

Han Sae Jung[1,‡], Johannes Cremer[2,‡], Aoyang Zhang[1,3,‡], Sangha Kim[1], Guang Yang[1], Ronald L. Walsworth[2,4,5,6,*], and Donhee Ham[1,*]

[1]John A. Paulson School of Engineering and Applied Sciences, Harvard University, Cambridge, MA 02138, USA.
[2]Quantum Technology Center, University of Maryland, College Park, MD 20742, USA
[3]School of Integrated Circuits, Tsinghua University, Haidian District, Beijing 100084, China
[4]Department of Physics, University of Maryland, College Park, MD 20742, USA
[5]Joint Quantum Institute, University of Maryland, College Park, MD 20742, USA
[6]Department of Electrical and Computer Engineering, University of Maryland, College Park, MD 20742, USA
[‡]These authors contributed equally to this work.
*Corresponding author. Email: donhee@seas.harvard.edu, walsworth@umd.edu

**This PDF file includes:**

    Methods

    Table S1

    Figures S1 to S8

    References

METHODS

**Chip packaging**

To package the chip, we first glue the chip's backside, which has a 10-μm thick gold layer, to the on-PCB gold plane via conductive silver epoxy; the on-PCB gold plane is internally connected to the on-PCB microstrip ground plane. Then, the two bond pads of the 3-turn loop are wire-bonded to the on-PCB microstrip signal lines. The diamond sample, which is custom-fabricated from Element Six, has NV centers embedded on one of its surfaces down to 10-μm depth, and is placed on top of the loop such that the loop is completely covered by the NV-embedded diamond surface. The diamond is glued to the chip by dropping PDMS on its edges and letting the PDMS cure at 80°C by placing the PCB on a hot plate for 15 min.

**Optically detected magnetic resonance (ODMR) measurement**

A microwave (MW) pulse of 20 μs duration is applied at the same time as an acousto-optic modulator (AOM) pulse (i.e., 532 nm green laser pulse) of 40 μs duration. A DAQ readout pulse of 1 μs duration is applied twice: 2 μs before the end of the MW pulse (i.e., at the 18 μs mark) and 2 μs before the end of the AOM pulse (i.e., at the 38 μs mark). The second readout, which happens in the absence of a MW pulse, serves as a reference readout; the spins are polarized at the $m_s = 0$ state, at which the red NV photoluminescence (PL) emission is at maximum. The PL contrast is calculated by comparing the signal readout (i.e., first readout) to the reference readout (i.e., second readout): $I_{contrast} = I_{sig} / I_{ref}$. This measurement is repeated for a range of different excitation frequencies of the MW pulse and averaged over 500 measurements. ODMR is observed when the PL contrast dips as the excitation frequency matches $f_0$.

To change $f_0$ as well as the DC magnet field $B_0$ alignment to one of the four NV axes, a permanent magnet attached to a motorized stage (Figure S1) is moved linearly in three dimensions while monitoring the ODMR spectrum. NV-centers within an ensemble are typically aligned in equal proportion along all four symmetry axes with the diamond crystal host. An external static (bias) $B_0$ field has a vector projection on all four axes, which results in multiple small PL contrast dips in the ODMR spectrum. A total of 4 × 2 = 8 PL dips are possible in the ODMR spectrum (with sub-structure to each PL dip arising from hyperfine splitting), corresponding to the projections of $B_0$ on the four NV axes and the $m_s = 0$ to $m_s = +1$ and $m_s = 0$ to $m_s = -1$ electronic spin resonance (ESR) transitions. When $B_0$ is aligned to one of the four NV axes, we observe one sharp large contrast PL dip for each ESR transition (corresponding to the NV axis aligned with $B_0$) and three small contrast PL dips for each ESR transition, corresponding to the vector projections of $B_0$ onto the remaining three NV axes.

**Rabi oscillation measurement**

A MW pulse of a varying duration is applied at $f_0$, determined from the ODMR measurement. An AOM pulse of 100 μs duration is applied 1 μs after the end of the MW pulse. A DAQ readout pulse of 1 μs duration is applied 0.5 μs after the start of the AOM pulse. The long AOM pulse duration remaining after the end of the DAQ pulse serves to re-initialize the NV spin to the $m_s = 0$ state before applying the next MW pulse of a different duration. The measurement is repeated for a range of different MW pulse durations and averaged over 500 measurements. How the two levels ($m_s = 0$ state and either the $m_s = +1$ state or the $m_s = -1$ state) are populated depends on the MW pulse duration, which controls the extent of the Rabi oscillation. This dependence is captured by the NV PL contrast plotted as a function of the MW pulse duration, showing the Rabi oscillation dynamics.

**Rabi frequency mapping**

To determine the microwave field $B_1$ homogeneity within the 3-turn loop, we map the NV ensemble Rabi frequency $f_1$, which is directly proportional to the $B_1$ magnitude. We measure $f_1$ across a 280 μm × 280 μm area surrounding the innermost turn of the 3-turn loop using a 5-μm diameter laser spot and a 10-μm pixel pitch. The loop's interior is scanned using a motorized stage by moving the chip stage with respect to the fixed laser spot location.

**Coherently averaged synchronized readout (CASR) detection of an RF signal**

To detect a $f_{signal} = 29.992$ MHz RF signal from a test coil using CASR, an XY8-6 sequence is applied to the NV ensemble with a central frequency $f_{casr} \approx 30$ MHz (with $\tau = 1/(2f_{casr})$ being the delay between π pulses). With a 2 μs delay after the XY8-6 sequence ends, an AOM pulse of green laser light of 20 μs duration is applied. With a 1.2 μs delay time after the AOM pulse begins, a DAQ readout pulse of 1 μs duration is applied to measure the NV PL. With a 1 μs delay time after the AOM pulse ends, the next XY8-6 sequence begins. This procedure is repeated until the measurement time reaches 1 s. The frequency of the measured NV PL variation, which is frequency-down-converted from the RF signal by the CASR protocol, is expected to be equal to the absolute difference between $f_{casr}$ and $f_{signal}$: $\delta f = f_{signal} - f_{casr} \approx |30$ MHz $- 29.992$ MHz$| \approx 8$ kHz. The Fourier transform of the measured CASR PL signal shows a single peak at this expected frequency, with a small difference within the 1 ppm error of the signal generator (Rigol DG 1032). The spectral resolution of a CASR measurement is given by the full width at half maximum of the signal peak in the Fourier transform. The Fourier limit to this spectral resolution is set by the reciprocal of the measurement time. CASR detection of the same RF signal is also performed for measurements times of 10 s and 100s, with spectral resolution given by the Fourier limit for the 1 s and 10 s measurements, and by about 0.05 Hz for the 100s measurement, limited by drift in the system (Figure S8).

| Work | Single NV vs. Ensemble | Location | ESR Frequency $f_0$ (GHz) | Rabi Frequency $f_1$ (MHz) | $f_1 / f_0$ | MW Power (W) | Driving Efficiency (MHz/$\sqrt{W}$) |
|---|---|---|---|---|---|---|---|
| This work | Ensemble | Center | 2.55 (optimal) | 136.3 | $5.3 \times 10^{-2}$ | 34.8 | 23.1 |
| Ref. 4 | Ensemble | Center | 2.87 | 14.3 | $5.0 \times 10^{-3}$ | 0.5 | 20.2 |
| Ref. 5 | Ensemble | Center | 2.82 | 4.6 | $1.6 \times 10^{-3}$ | 1.0 | 4.6 |
| Ref. 7 | Ensemble | Center | 2.67 | 7.9 | $3.0 \times 10^{-3}$ | 4.0 | 3.9 |
| Ref. 3 | Single | Close to trace | 0.49 | 440 | $9.0 \times 10^{-1}$ | 0.4 | 697.4 |
| Ref. 6 | Ensemble | At trace | 2.73 | 165 | $6.0 \times 10^{-2}$ | 3.4 | 89.6 |

**Table S1.** Comparison of device used in present work to several other works in the literature, both those that measure NV Rabi oscillations at the center of a loop (lower Rabi frequency and better microwave field spatial homogeneity) and directly at or near the metal trace (higher Rabi frequency and poorer homogeneity).

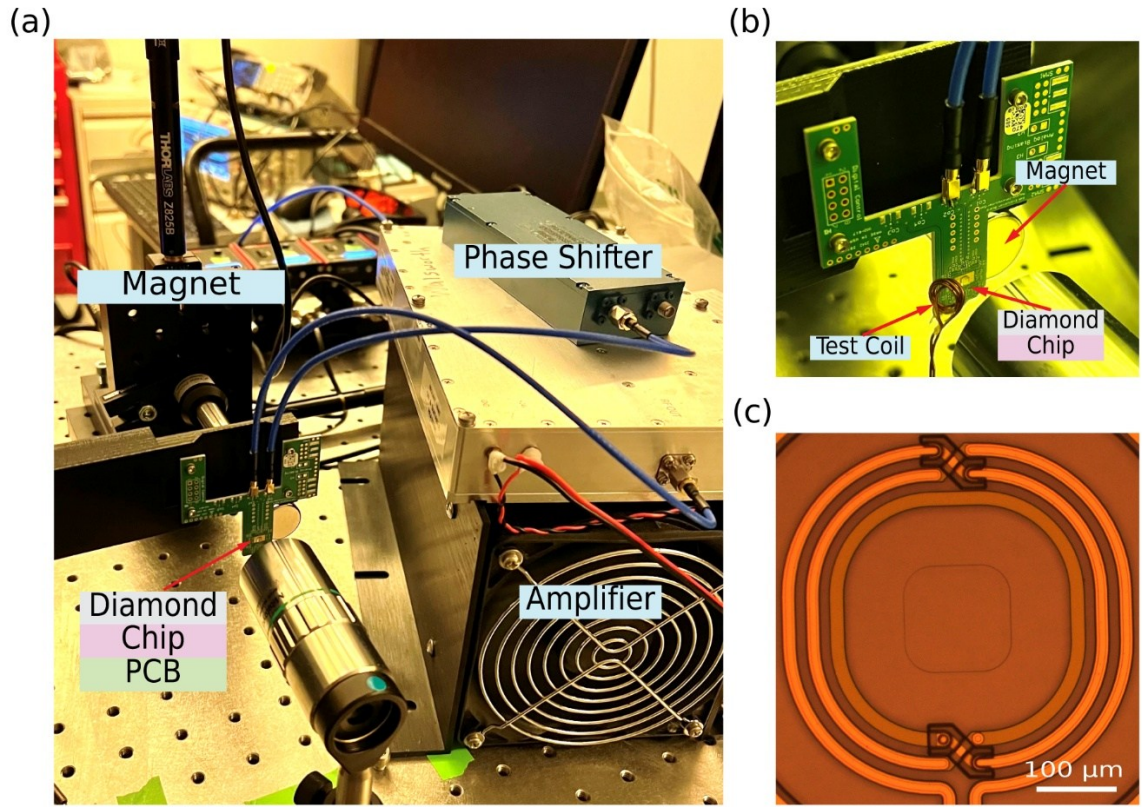

**Figure S1.** Experimental setup. (a) Photograph of setup. The 3-turn wire loop is fabricated on a semi-insulating SiC substrate, which is glued to the PCB's gold ground plane via conductive epoxy. Diamond is glued on top of the loop using PDMS, with the nitrogen-vacancy (NV) layer facing the loop. (b) Close-up view of the setup during a coherently averaged synchronized readout (CASR) detection of an RF signal from a test coil. (c) Optical micrograph of the 3-turn loop. The metal trace is 17 μm wide, with the gap between the metal traces being 13 μm. The diameters (center-to-center distance between the metal traces) of the innermost loop, the second loop, and the outer-most loop are 300 μm, 360 μm, and 420 μm, respectively.

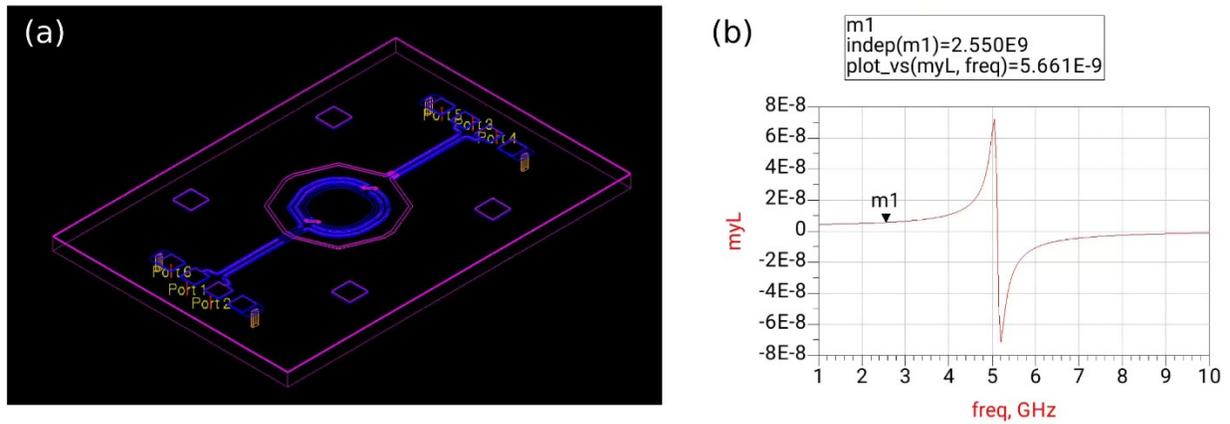

**Figure S2.** Electromagnetic (EM) simulation of the 3-turn loop. (a) EM model of the passive chip in ADS Momentum. (b) EM simulation of the 3-turn loop inductance as a function of microwave excitation frequency yields inductance of 5.7 nH at 2.55 GHz.

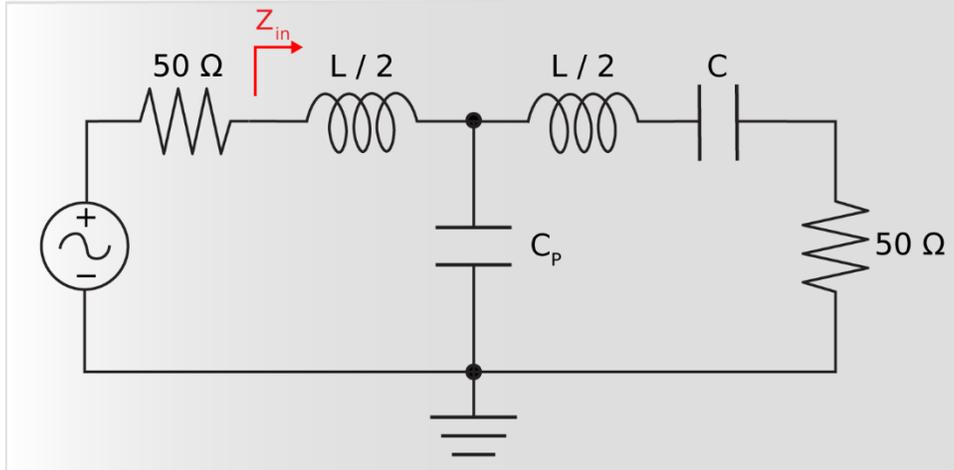

**Figure S3.** Circuit model of the microwave signal pathway using the 50-Ω termination. Instead of using an open-ended phase shifter termination (Figure 1c), the microwave pathway is terminated with a 50-Ω resistor connected to ground.

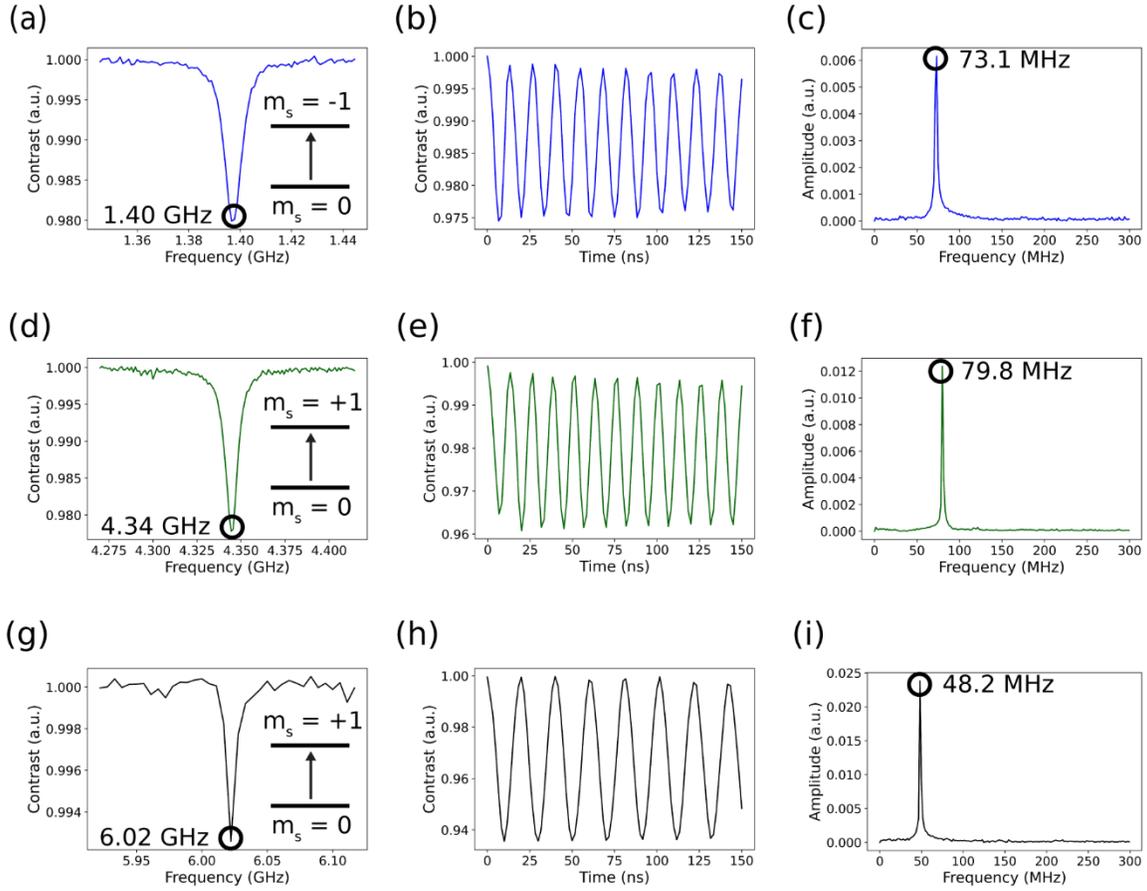

**Figure S4.** NV ODMR and Rabi oscillation measurements at various $f_0$ values. (a) ODMR spectrum for the $m_s = 0$ to $m_s = -1$ NV ESR transition at $f_0 = 1.40$ GHz ($B_0 = 526$ G). (b) Rabi oscillation measurement using the ESR transition in (a). (c) Fourier transform of the Rabi oscillation measurement in (b), showing a single peak at $f_1 = 73.1$ MHz. (d) ODMR spectrum of the $m_s = 0$ to $m_s = +1$ NV ESR transition at $f_0 = 4.34$ GHz ($B_0 = 526$ G). (e) Rabi oscillation measurement using the ESR transition in (d). (f) Fourier transform of the Rabi oscillation measurement in (e), showing a single peak at $f_1 = 79.8$ MHz. (g) ODMR spectrum of the $m_s = 0$ to $m_s = +1$ NV ESR transition at $f_0 = 6.02$ GHz ($B_0 = 1125$ G). (h) Rabi oscillation measurement using the ESR transition in (g). (i) Fourier transform of the Rabi oscillation trace measurement in (h), showing a single peak at $f_1 = 48.2$ MHz.

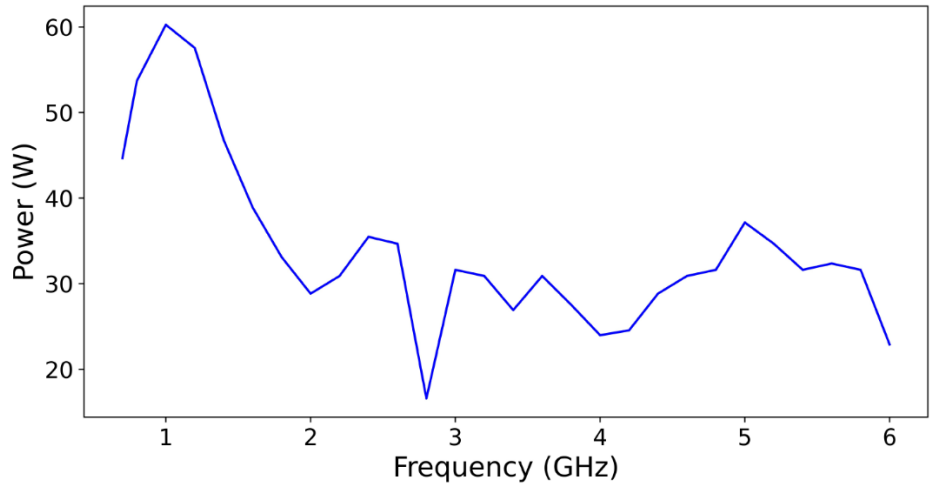

**Figure S5.** Output power of the Mini-Circuits ZHL-25W-63+ power amplifier as a function of microwave excitation frequency, derived from the vendor's publicly available data set describing the amplifier's saturated power output (shown in dBm in their document) as a function of excitation frequency (see Reference 1 of the Supporting Information).

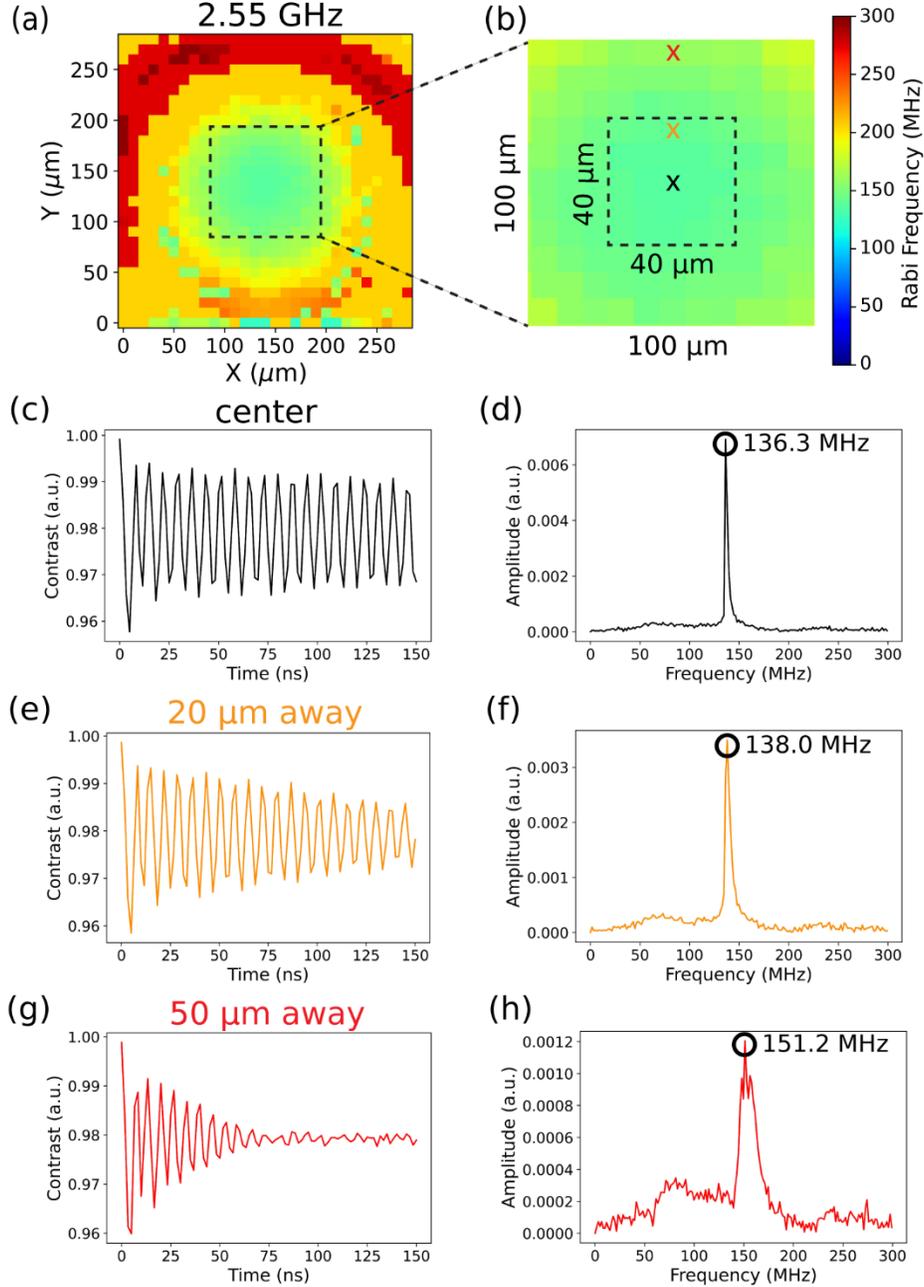

**Figure S6.** $f_1$ measurements at varying distances from the center of the 3-turn loop for the NV-measured map of $f_1$ shown in Figure 3, left at $f_0$ = 2.55 GHz. (a) $f_1$ map shown in Figure 3, left. The 100 μm × 100 μm area around the center is denoted by the dashed square. (b) Close-up view of the 100 μm × 100 μm area shown in (a). The 40 μm × 40 μm area around the center is denoted by the dashed square. The black, yellow, and red crosses denote the locations that are 0 μm, 20 μm, and 50 μm away from the center, respectively. (c) Rabi oscillation measurement at the center. (d) Fourier transform of the Rabi oscillation measurement in (c), showing a single peak at $f_1$ = 136.3 MHz. (e) Rabi oscillation measurement at the 20 μm mark. (f) Fourier transform of the Rabi oscillation measurement in (e), showing a single peak at $f_1$ = 138.0 MHz. (g) Rabi oscillation measurement at the 50 μm mark. (h) Fourier transform of the Rabi oscillation measurement in (g),

showing a peak at $f_1$ = 151.2 MHz. Moving away from the loop center, there is increasing spatial variation in microwave field strength, leading to enhanced damping of observed Rabi oscillations and corresponding broadening of the Fourier transform.

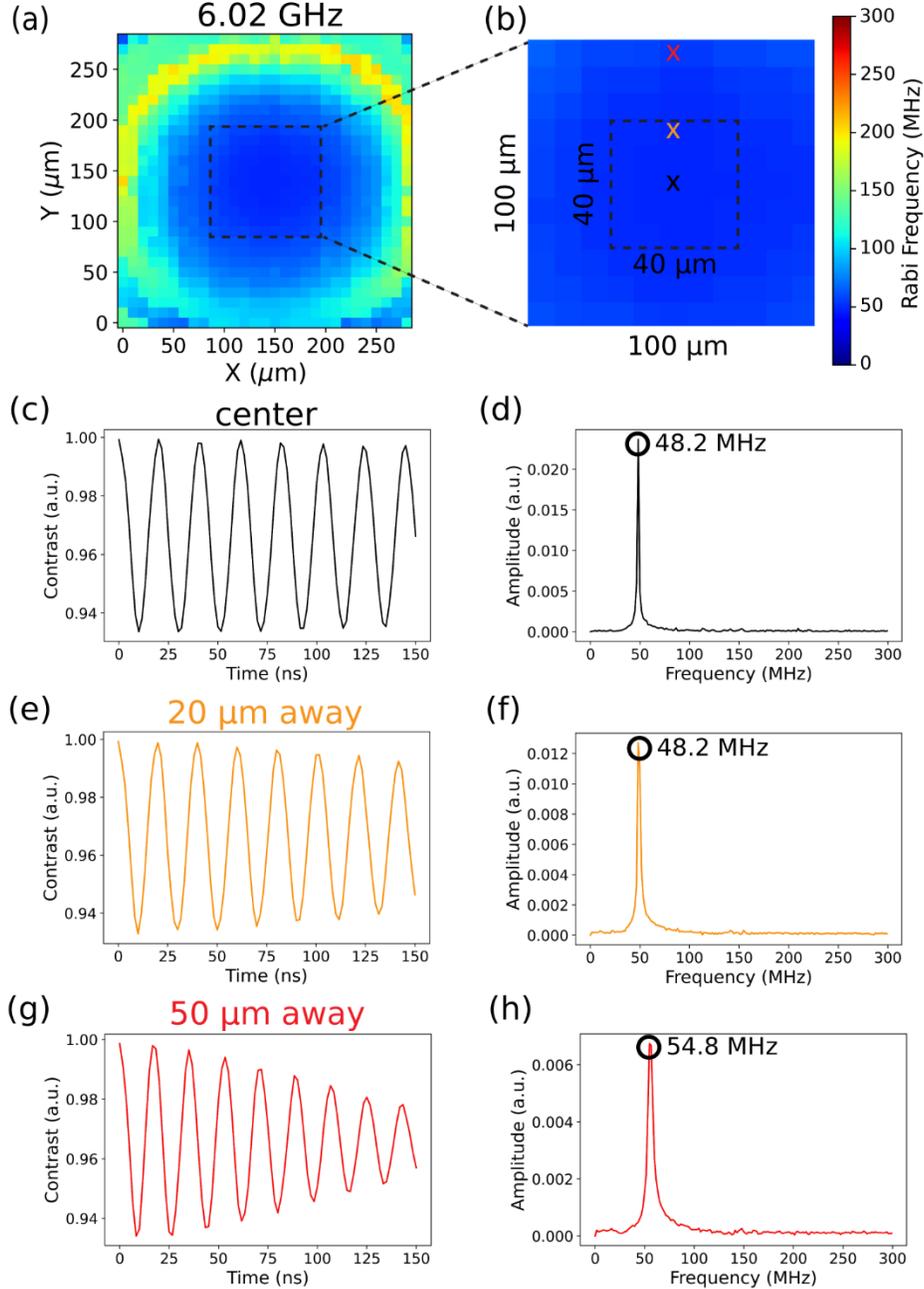

**Figure S7.** $f_1$ measurements at varying distances from the center of the 3-turn loop for the NV-measured map of $f_1$ shown in Figure 3, right at $f_0 = 6.02$ GHz. (a) $f_1$ map shown in Figure 3, right. The 100 μm × 100 μm area around the center is denoted by the dashed square. (b) Close-up view of the 100 μm × 100 μm area shown in (a). The 40 μm × 40 μm area around the center is denoted by the dashed square. The black, yellow, and red crosses denote the locations that are 0 μm, 20 μm, and 50 μm away from the center, respectively. (c) Rabi oscillation measurement at the center. (d) Fourier transform of the Rabi oscillation measurement in (c), showing a single peak at $f_1 = 48.2$ MHz. (e) Rabi oscillation measurement at the 20 μm mark. (f) Fourier transform of the Rabi oscillation measurement in (e), showing a single peak at $f_1 = 48.2$ MHz. (g) Rabi oscillation measurement at the 50 μm mark. (h) Fourier transform of the Rabi oscillation measurement in (g),

showing a peak at $f_1$ = 54.8 MHz. As in Figure S6, there is increasing spatial variation in microwave field strength moving away from the loop center, leading to enhanced damping of observed Rabi oscillations and corresponding broadening of the Fourier transform.

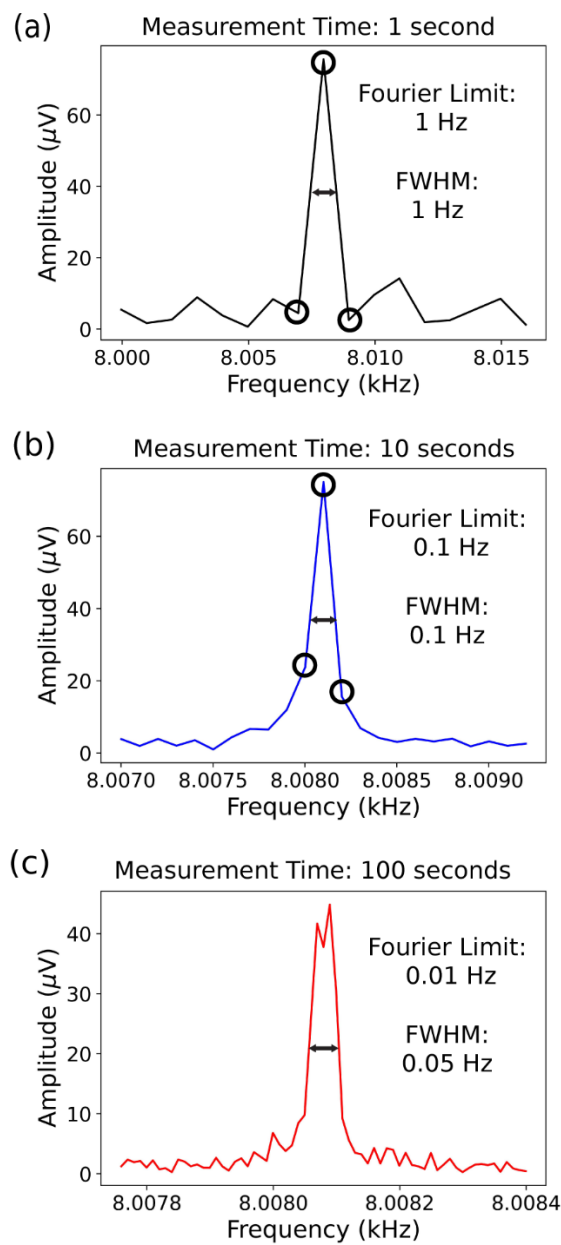

**Figure S8.** Determination of the spectral resolution for the coherently averaged synchronized readout (CASR) measurement shown in Figure 4 at varying measurement times. (a) Fourier transform for a 1 s CASR measurement. The spectral resolution, determined by the full width at half maximum (FWHM) of the Fourier transform, is set by the Fourier limit, i.e., 1 Hz for a 1 s measurement. (b) Fourier transform for a 10 s CASR measurement. The spectral resolution is again set by the Fourier limit (0.1 Hz). (c) Fourier transform for a 100 s CASR measurement. The spectral resolution is about 0.05 Hz, limited by system drift.